# Root Cause Analysis and Correction of Single Metal Contact Open-Induced Scan Chain Failure in 90nm node VLSI

*Chao-Cheng Ting[ab]\*, Ya-Chi, Liu[c], Hsuan-Hsien Chen[a], Chung-Ching Tsai[a] and Liwen Shih[d]*
[a]Division of IC Technology, Faraday Technology Corporation, No.5, Li-Hsin Rd III, Hsin Chu Science park, Hsin Chu, Taiwan 300, R.O.C
[b]Department of Materials science and Engineering, National Chiao-Tung University, 1001 Ta-Hsueh Road, Hsinchu, 30010 Taiwan, R.O.C
[c]Optmic Lab, 340 S. Lemon Ave. #8365, Walnut, CA 91789, U.S.A.
[d]Department of Computational Engineering, University of Houston-Clear Lake, Houston, U.S.A.
*Corresponding author:
Department of Materials Science and Engineering, National Chiao-Tung University, 1001 Ta Hsueh Road, Hsinchu, Taiwan 30010, R.O.C.

*Corresponding author:*
E-mail: Chaochengting.MSE00g@g2.nctu.edu.tw

## Abstract

In this paper, the localization of open metal contact for 90nm node SOC is reported based on Electron Beam Absorbed Current (EBAC) technique and scan diagnosis for the first time. According to the detected excess carbon, silicon and oxygen signals obtained from X-ray energy dispersive spectroscopy (EDX), the failure was deemed to be caused by the incomplete removal of silicate photoresist polymer formed during the $O_2$ plasma dry clean before copper plating. Based on this, we proposed to replace the dry clean with diluted HF clean prior to the copper plating, which can significantly remove the silicate polymers and increase the yield.

## Introduction

As transistors shrink in size, the localization of integrated circuits (IC) failures is facing a great challenge to detect weak leakage-current-induced thermal emissions [1]. The challenge is aggravated when metal voids that occurred in very-large-scale-integration (VLSI) do not give rise to thermal emissions, hindering emission microscopy (EMMI) detection. Electron Beam Absorbed Current (EBAC) technique uses an electron beam as a probing element to achieve nano-meter scale resolution in this isolation of open failures in VLSI [1, 2] based on the repulsion nature of the electron signal at the open spot. In this paper, we propose a combination of scan diagnosis and EBAC to locate open defects with faster turnaround and precision.

First, scan diagnosis is used to narrow down the suspected failure loop before EBAC locates the failing path from the loop. Focused Ion Beam (FIB) cross section image reveals the copper void causing the open contact. Finally, Transmission Electron Microscope (TEM) and X-ray energy dispersive spectroscopy (EDX) analysis are conducted to identify the failure mechanism. The investigation results show that the failure was possibly due to the incomplete removal of photo-resist residues and solvent at the via bottom that results in the formation of silicate polymer during the plasma dry clean prior to copper plating. To substantiate this hypothesis, the fabrication flow was modified with an extra diluted HF cleaning prior to copper plating to replace the dry cleaning. With the modified flow, it was found that the silicate polymer is significantly removed and the yield greatly enhanced from approximately 34% to 90%.

## Discussion

### Background of the problem

A 90 nm node VLSI based on standard performance process suffered significant low yield of approximately 34%. The failing wafer map is shown in Figure 1. With reference to Figure 1, the two top failures are identified as scan chain test and ATPG (Automatic Test Pattern Generation) test. Since the scan chain failure dominates, this paper studies the cause and corrective action to recover the scan chain failure.

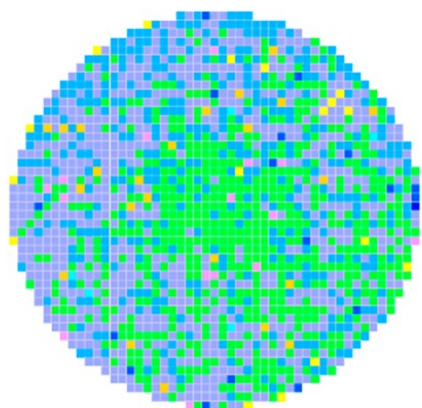

| 34(%) | Pass die rate |
| --- | --- |
| 40(%) | Fail rate of scan test |
| 22(%) | Fail rate of ATPG test |

**Figure 1.** The failing wafer map. (Green) Passing dies. (Purple) Dies failing scan chain. (Blue) Dies failing ATPG test.

### Scan Diagnosis and Sample preparation

First, scan diagnosis was performed to localize the suspected failure net as shown in Figure 2 (top). Based on the scan diagnosis result, the sample was polished to expose via 5 for probing. As shown in Figure 2 (bottom), the power was supplied to via 5 and the ground was applied to the substrate for further EBAC analysis.

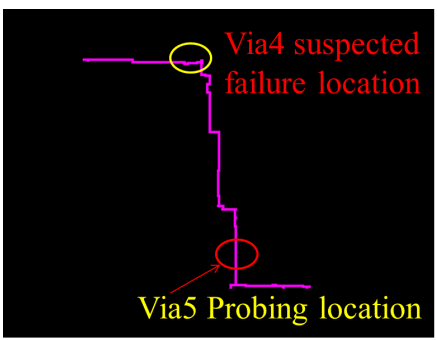

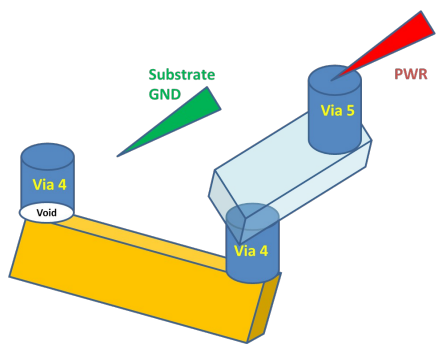

**Figure 2.** The scan diagnosis indicating the suspected failure loop (top) and the illustration of probing for EBAC imaging (bottom).

### Fault isolation using EBAC

EBAC technique was performed on that suspected trace to identify the failure location. Figure 3 shows the EBAC image indicating the failure location to be around via 4 as shown by the disconnected metal trace after via 4 as compared to the scan diagnosis suspected net; this technique provides a quick way to assess the suspected failing net without physically inspecting the entire net which can be time-consuming.

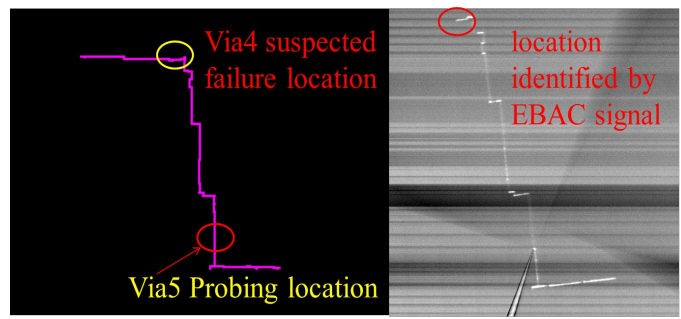

**Figure 3.** The scan diagnosis result of failure trace (left) and the EBAC image (right).

### Material and Root Cause analysis

Figure 4 shows the SEM image of the FIB cross section revealing a sub-surface void at the bottom of via 4, which is of the nature as reported by Richard L., et al [3]. In addition, there is no deformed via as compared to the adjacent ones. This indicates that the open failure does not belong to the via etch or former lithography process.

In order to uncover the actual cause of the open failure, we performed EDX element analysis. As shown in Figure 5(c) and (f), the extra carbon and oxygen signals were discovered at the bottom of the metal contact. Based on the above information and the presence of silicon signal, as shown in Figure 5(b), we reasonably assumed that the Si, O and C signals come from silicate polymer as it cannot be removed by wet solvent cleaning as depicted in Figure 6.

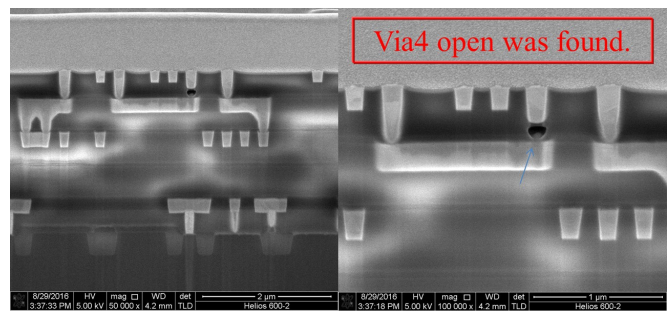

**Figure 4.** SEM overview image of the FIB cross section (left) and the high magnitude SEM image indicating via 4 open (right).

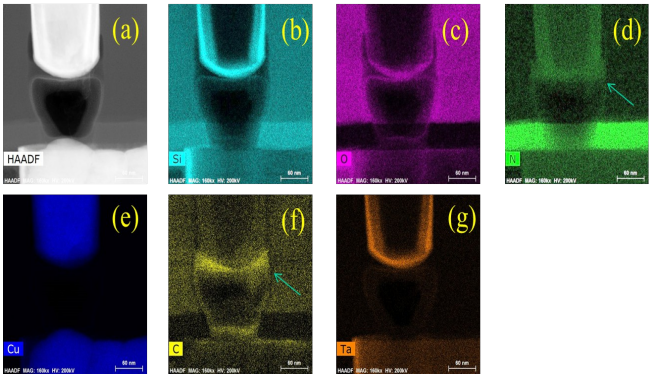

**Figure 5.** TEM and EDX mapping image. (a) The HAADF (High Angle Annular. Dark Field) TEM image. (b) The EDX mapping of Si (light blue). (c) The EDX mapping of O (pink). (d) The EDX mapping of N (green). (e) The EDX mapping of Cu (dark blue). (f) The EDX mapping of C (yellow). (g) The EDX mapping of Ta (brown).

### Process flow and partition check analysis

A partition check analysis was conducted using defect performance inspection at each process step. It was found that the failure likely occur in the steps after the first solvent clean, as indicated in Figure 6. Based on this insight, we postulate that the formation of silicate polymer was due to the reaction of the silicon and photoresist radicals generated during the plasma dry clean process, as indicated by the red spot in Figure 6. This is because unlike the organic compounds (photoresist), the silicate polymer cannot be easily removed during the wet clean. Also, based on Figure 6, the solvent may remain in the silicate polymer and outgas during the pre-barrier baking process, which can lead to poor copper adhesion in the following electrochemical plating process (Koh, L. T., et al argued the carbon containment may affect the bottom fill up ability and He, Q. Y et al noted the residual carbon contamination will increase the occurrences of copper void during the ECP. [4,-6]). To substantiate our hypothesis, we devised a revised fabrication flow with ultra-dilute hydrogen fluoride (UDHF) clean and $N_2$ purge to replace the dry clean and second solvent clean to avoid radical recombination and solvent outgassing during pre-barrier bake respectively.

As shown in Figure 7(a), the UDHF clean was applied just after the solvent clean and before the DI wafer flush to ensure the total removal of the expected silicate polymer. The scan chain fail rate was significantly reduced with the revised fabrication flow. The wafermap with the improved yield is presented in Figure 7(b). The yield improvement from approximately 34% to 90% as a result of the corrective action directly supports our assumption on the formation of silicate polymer.

For a complete understanding on the failure mechanism, Figure 8 illustrates the formation of silicate polymer in three stages. First, after the etching process, the photoresist and solvent residue that remain at the bottom of via are hard to be completely removed by the wet clean process. Second, the $O_2$ plasma dry clean is capable of ionizing the photoresist and solvent residue, leading to the formation of Si, C and O radicals. Third, after the plasma dry clean, the Si, C and O radicals recombined to form silicate polymer which give rise to C, Si, O signals in the EDX images. Finally, the unwanted silicate polymer along with the solvent residue evaporate during the later pre-barrrier baking process leading to a poor adhesion in electrochemical plating process. This explains the subsequent copper void.

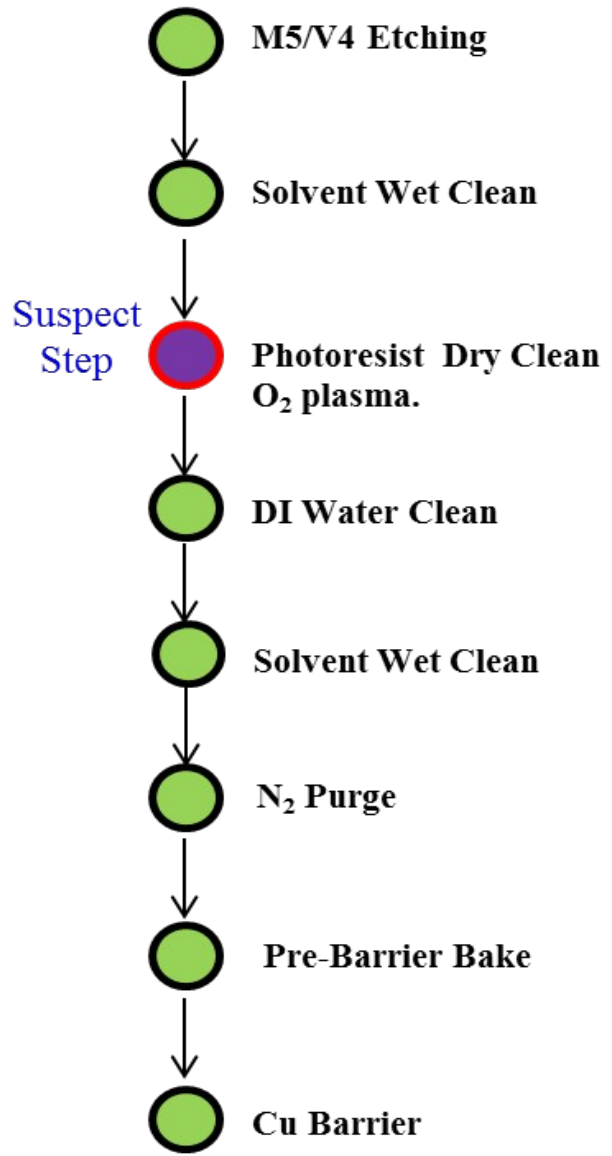

**Figure 6.** The fabrication flow and the suspected photoresist $O_2$ plasma dry clean step.

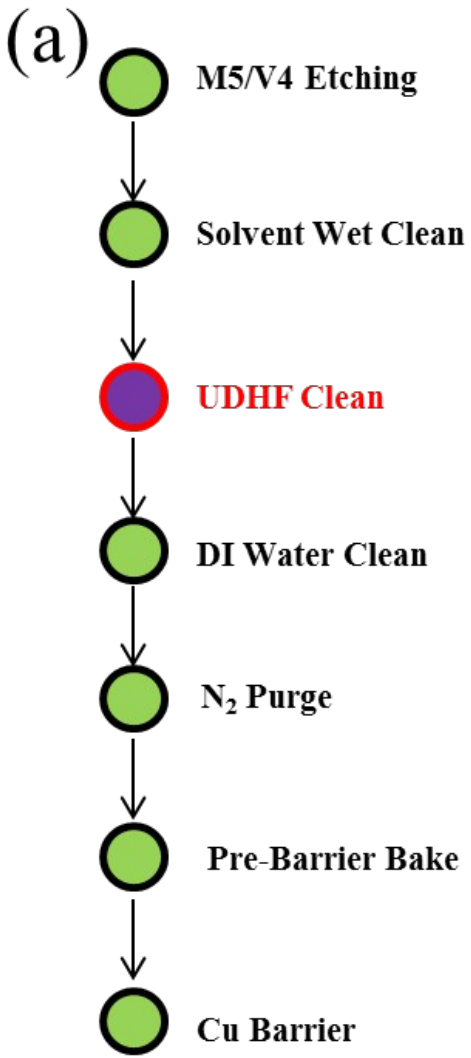

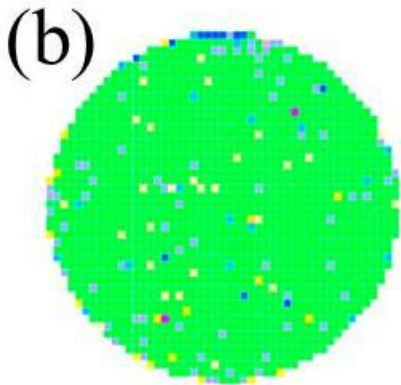

**Figure 7**. (a) The revised fabrication flow. (b) The test result of the revised flow. (Green) Passing dies on all test items.

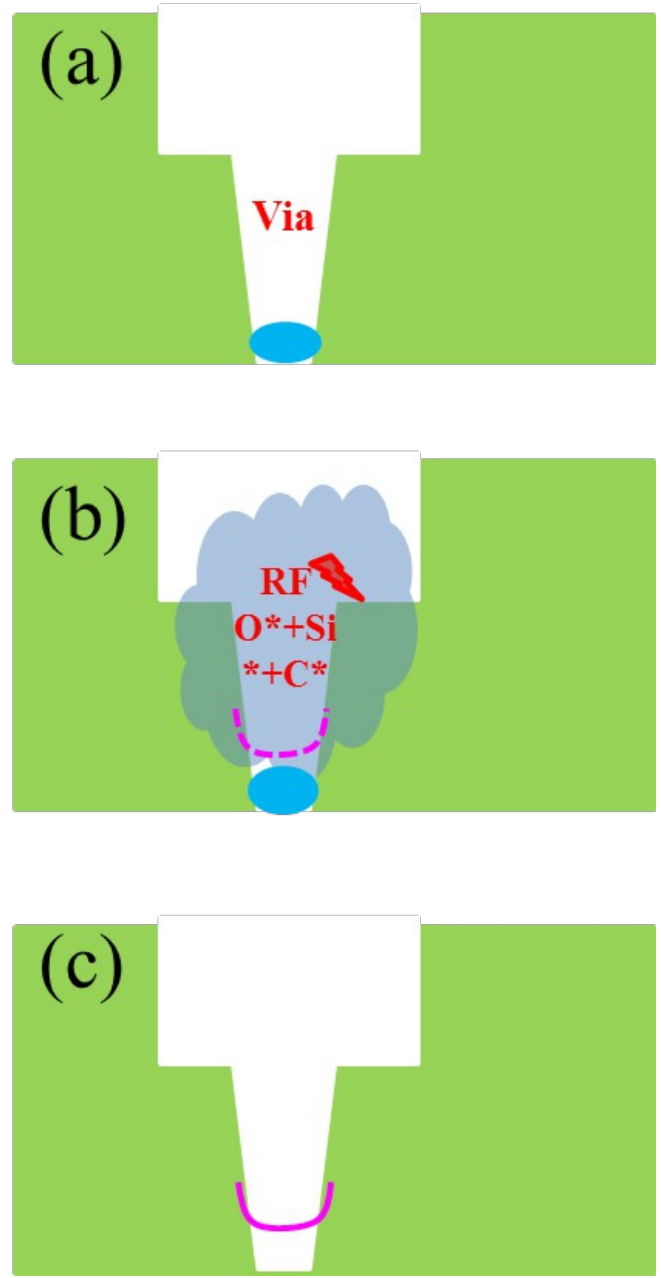

**Figure 8.** The schematic of the failure mechanism. (a) solvent and photo-resist residue (carbonaceous compound) adhere to the via bottom. (b) The $O_2$ plasma dry clean.generate O, Si and C radicals. (c) The unwanted silicate polymer is formed at the via bottom based on the O, Si and C radical recombination .

## Conclusions

This study demonstrates the effectiveness of using a combination of scan diagnosis and EBAC to locate the metal contact failure for complex VLSI in 90nm technology node.

Subsequent EDX analysis at the failing location reveal the root cause of the open metal contact to be due to the deposition of silicate polymer formed by the recombination of O, Si and C radicals. To fix this problem, we modified the fabrication process by implementing an extra UDHF cleaning and reducing the oxygen plasma processes to prevent possible formation of the undesirable O, Si and C radicals during the plasma dry clean. The yield enhancement successfully proves the failure mechanism.

## Acknowledgments

We acknowledge the failure analysis support of United Microelectronics Corporation and Materials Analysis Technology Inc.